\documentclass[fleqn,usenatbib,twocolumn,beamer]{mnras}

\usepackage{newtxtext,newtxmath}

\usepackage[T1]{fontenc}

\DeclareRobustCommand{\VAN}[3]{#2}
\let\VANthebibliography\thebibliography
\def\thebibliography{\DeclareRobustCommand{\VAN}[3]{##3}\VANthebibliography}

\usepackage{subfig,graphicx}	
\usepackage{amsmath}	
\usepackage{booktabs}
\usepackage{float}
\usepackage{placeins}
\usepackage{outlines}
\usepackage{enumitem}
\usepackage[english]{babel}
\usepackage{blindtext}
\usepackage{tabularx}

\usepackage{xcolor, soul}

\title[Radio Relic in Abell 4067]{Discovery of a nearby radio relic in the low-mass, merging cluster Abell~4067}

\author[Magolego et al.]{
Isaac Magolego$^{1}$\thanks{E-mail: isaacike07@gmail.com},
Roger P. Deane$^{1,2,3}$,
Kshitij Thorat$^{3}$, Ian Heywood$^{4,5,6,7}$, Justin Spilker$^{8}$,
\newauthor
\hspace*{0.2em}Taweewat Somboonpanyakul$^{9}$, Dazhi Zhou$^{10}$, Manuel Aravena$^{11,12}$, Joaquin D.~Vieira$^{13,14}$,
\newauthor
\hspace*{0.2em}Kedar A.~Phadke$^{13,14,15}$, Lindsey E.~Bleem$^{16,17,18}$, Scott C.\ Chapman$^{19,10,20}$
\\\\
$^{1}$Wits Centre for Astrophysics, School of Physics, University of the Witwatersrand, Private Bag 3, 2050, Johannesburg, South Africa\\
$^{2}$Inter-University Institute for Data Intensive Astronomy, Department of Astronomy, University of Cape Town, Cape Town, South Africa\\
$^{3}$Department of Physics, University of Pretoria, Hatfield, Pretoria, 0028, South Africa\\
$^{4}$SKA Observatory, Jodrell Bank, Lower Withington, Macclesfield, SK11 9FT, UK\\
$^{5}$Astrophysics, Department of Physics, University of Oxford, Keble Road, Oxford, OX1 3RH, UK\\
$^{6}$Department of Physics and Electronics, Rhodes University, PO Box 94, Makhanda 6140, South Africa\\
$^{7}$South African Radio Astronomy Observatory, Liesbeek House, River Park, Gloucester Road, Mowbray, 7700, South Africa\\
$^{8}$Department of Physics and Astronomy and George P. and Cynthia Woods Mitchell Institute for Fundamental Physics and Astronomy,\\
Texas A\&M University, 4242 TAMU, College Station, TX 77843-4242, USA\\
$^{9}$Department of Physics, Faculty of Science, Chulalongkorn University, 254 Phayathai Road, Pathumwan, Bangkok 10330, Thailand\\
$^{10}$Department of Physics and Astronomy, University of British Columbia, 6225 Agricultural Road, Vancouver V6T 1Z1, Canada\\
$^{11}$Instituto de Estudios Astrof\'{\i}cos, Facultad de Ingenier\'{\i}a y Ciencias, Universidad Diego Portales, Av. Ej\'ercito 441, Santiago, Chile\\
$^{12}$Millenium Nucleus for Galaxies (MINGAL)\\
$^{13}$Department of Astronomy, University of Illinois, 1002 West Green Street, Urbana, IL 61801, USA\\
$^{14}$Center for AstroPhysical Surveys, National Center for Supercomputing Applications, Urbana, IL, 61801, USA\\
$^{15}$NSF-Simons AI Institute for the Sky (SkAI), 172 E. Chestnut St., Chicago, IL 60611, USA\\
$^{16}$High-Energy Physics Division, Argonne National Laboratory, 9700 South Cass Avenue., Lemont, IL, 60439, USA\\
$^{17}$Kavli Institute for Cosmological Physics, University of Chicago, 5640 South Ellis Avenue, Chicago, IL, 60637, USA\\
$^{18}$Department of Astronomy and Astrophysics, University of Chicago, 5640 South Ellis Avenue, Chicago, IL, 60637, USA\\
$^{19}$Department of Physics and Atmospheric Science, Dalhousie University, 6310 Coburg Road, Halifax B3H 4R2, Canada\\
$^{20}$Eureka Scientific Inc, Oakland, CA 94602, USA\\}

\date{Accepted XXX. Received YYY; in original form ZZZ}

\pubyear{2015}

\begin{document}
\label{firstpage}
\pagerange{\pageref{firstpage}--\pageref{lastpage}}
\maketitle

\begin{abstract}
Shock waves generated during cluster mergers offer a powerful probe of how large-scale structure grows and evolves in the Universe. As part of the MeerKAT-South Pole Telescope (SPT) survey, we report the discovery of a single arc-like radio relic in the galaxy cluster Abell~4067 ($z=0.099$), one of the lowest-mass clusters known to host such a structure. MeerKAT UHF-band (0.58 - 1.09 GHz) observations reveal a relic with a largest linear size of $\sim1.48\pm0.02$~Mpc, located at a projected distance of 0.95~Mpc from the cluster centre. XMM-Newton X-ray data show that the relic’s position and orientation relative to the intracluster medium (ICM) elongation are consistent with a merger-driven shock-wave scenario. The relic has an estimated radio power of $3.10\pm0.03\times10^{24}$~W~Hz$^{-1}$ at 150~MHz. When placed in the $P_{150\,\mathrm{MHz}}$-$M_{500}$ scaling relation, the Abell 4067 relic appears less luminous compared to relics in more massive clusters, suggesting an association with weak merger shocks. This finding supports the idea that relics in low-mass clusters may form through less energetic merger events, leading to weak merger shocks. The latter is supported by the absence of a detectable central radio halo in Abell~4067, which reinforces the idea that luminous radio halos are not a universal outcome of cluster mergers and highlights the role of cluster mass, merger energetics and evolutionary stage in shaping diffuse radio emission in the intracluster medium.

\end{abstract}




\section{Introduction} 
Galaxy clusters undergoing mergers often exhibit large-scale, diffuse radio emission in the form of radio halos and radio relics (see \citeauthor{Weeren_2019} \citeyear{Weeren_2019} for a review). Radio relics are elongated synchrotron-emitting structures found at the peripheries of galaxy clusters, often appearing as striking single or double arcs (e.g., \citeauthor{Bagchi_2006} \citeyear{Bagchi_2006}; \citeauthor{Feretti_2012} \citeyear{Feretti_2012}; \citeauthor{Weeren_2019} \citeyear{Weeren_2019}). Unlike radio halos, which are associated with turbulence in the intracluster medium (ICM), relics are linked to merger-driven shock waves propagating through the ICM \citep{Sarazin_2013,Ogrean_2013,Botteon_2016a,Botteon_2016b}. These shocks play a crucial role in accelerating cosmic ray electrons (CRe), leading to the observed synchrotron emission \citep{Roettiger_1999,Hoeft_2007}.

A defining characteristic of radio relics is their steep radio spectra ($\alpha$ > 1, where $S_{v}$ $\propto$ $v^{-\alpha}$), which typically steepen in the direction towards the cluster centre, consistent with the energy losses of relativistic electrons (e.g., \citeauthor{Weeren_2019} \citeyear{Weeren_2019}). Additionally, radio relics exhibit high levels of polarization, with fractions of 20-30 percent at 1.4 GHz and up to $\sim$ 70 percent at 5 GHz, indicative of ordered magnetic fields aligned with the shock front \citep{Weeren_2010,Kierdorf_2017,Loi_2017}. The detection of X-ray brightness discontinuities at the location of several relics further supports their association with merger shock waves \citep{Akamatsu_2013,Botteon_2018}.

The standard framework for explaining radio relics invokes Diffusive Shock Acceleration (DSA; \citeauthor{Bell_1978} \citeyear{Bell_1978}; \citeauthor{Jones_1991} \citeyear{Jones_1991}; \citeauthor{Ensslin_1997} \citeyear{Ensslin_1997}), wherein electrons are accelerated at shock fronts via a first-order Fermi process. However, observations suggest that the weak shocks typically found in clusters (Mach numbers $\mathcal{M}$ $\sim$ 2) may not efficiently accelerate thermal electrons to relativistic energies \citep{Markevitch_2010,Botteon_2020a,Wittor_2021,Whittingham_2024}. Additionally, these shocks should also accelerate protons, which would produce $\gamma$-ray emission through interactions with the ICM, yet no such emission has been detected \citep{Ackermann_2016}, placing stringent limits on the acceleration efficiency of protons \citep{Vazza_2016}.

To address these challenges, alternative mechanisms have been proposed. One possibility is that shocks re-accelerate a pre-existing population of mildly relativistic electrons rather than accelerating thermal particles \citep{Markevitch_2005,Pinzke_2013,Kang_2015}. These seed electrons could originate from previous AGN activity \citep{Bonafede_2014,Weeren_2017,Stuardi_2019} or past merger events, making re-acceleration a more efficient process. Another proposed solution is Shock Drift Acceleration (SDA), a pre-heating mechanism in which particles gain energy by drifting along the shock front in the presence of magnetic field gradients and the motional electric field, thereby increasing the efficiency of subsequent DSA \citep{Matsukiyo_2011,Guo_2014}. 

Despite these theoretical advancements, several open questions remain regarding the injection and acceleration of electrons, the role of magnetic fields, and the overall energy budget required to sustain radio relics \citep{Brunetti_2014}. Additionally, the observed scaling relation between relic radio power and host cluster mass \citep{deGasperin_2014,Kale_2017,Jones_2023} hints at a deeper connection between cluster mergers and relic formation, though the scatter in single-relic systems suggests that additional factors influence their occurrence. A comprehensive analysis of a larger sample of relics is necessary to further our understanding of these enigmatic structures and their connection to large-scale structure formation.

Only a small ($\sim$ 10 percent) fraction of clusters are expected to host radio relics \citep[e.g.,][]{Kale_2015,Zhou_2022,Jones_2023,Lee_2024}, despite theoretical expectations and cosmological simulations that predict many more. To our knowledge, only $\sim$ 4 single-radio-relic systems have been reported in low mass clusters \citep[$M_{500} \lesssim 3 \times 10^{14} \, M_\odot$;][]{Kale_2017,Paul_2020,Paul_2021}. This scarcity highlights the importance of deep, high-sensitivity observations capable of detecting low-surface-brightness relic emission in less extreme environments. Future high-sensitivity, wide-frequency surveys with uniform $uv$-coverage and improved detection of low-surface brightness, steep-spectrum emission will thus be essential to build a far larger, statistically meaningful relic sample spanning a broad range of cluster masses and redshifts.

In this Letter, we report the discovery of a nearby radio relic, detected in MeerKAT UHF-band observations of the galaxy cluster RXCJ2359.3$-$6042 (Abell 4067), as part of the MeerKAT-SPT survey. This cluster, listed in the REFLEX II cluster sample \citep{REFLEXII_2013,Xu_2022}, is at a redshift of $z = 0.0994 \pm 0.0050$ \citep{Xu_2022} and has a mass of $M_{500} = 2.0 \pm0.25\times 10^{14}$~M$_{\odot}$ \citep{Chon_2015}, consistent within uncertainties with the SZ-derived value from the \textit{Planck} SZ catalogue, $M_{500} = 2.32 \times 10^{14}$~M$_{\odot}$ \citep{Tarr_2018}. The Letter is structured as follows: Section  \ref{sec2} outlines the observations and data processing, Section \ref{sec3} presents the results, Section \ref{sec4} discusses the results, and Section~\ref{sec5} offers our conclusions. We adopt a $\Lambda$CDM flat cosmology with $H_{0} = 70~\mathrm{km}~\mathrm{s}^{-1}~\mathrm{Mpc}^{-1}$, $\Omega_{m} = 0.3$, and $\Omega_{\Lambda} = 0.7$. At the redshift of Abell 4067 ($z = 0.0994$), 1 arcsec corresponds to 1.83 kpc.

\begin{table}
    \centering
    \caption{Properties of Abell 4067. X-ray properties are adopted from \citet{Chon_2015}.}
    \label{properties}
    \begin{tabular}{lc}
        \hline
        R.A.$_{\rm J2000}$ ($^{\mathrm{h}}$:$^{\mathrm{m}}$:$^{\mathrm{s}}$) & 23:58:49.9 \\
        Dec.$_{\rm J2000}$ ($^{\circ}\!:\!^{\prime}\!:\!^{\prime\prime}$) & $-$60:37:25 \\
        Redshift ($z$) & 0.0994 \\
        $M_{500c}$ (10$^{14}$~M$_{\odot}$) & 2.0 $\pm$ 0.25 \\
        $R_{500}$ (kpc) & 800 \\
        \hline
    \end{tabular}
\end{table}

\section{Data Processing and Product}
\label{sec2}
\subsection{Radio Data}
The 100 deg$^2$ MeerKAT-South Pole Telescope Survey  (Proposal ID:SCI-20220822-JV-02 and SCI-20230907-JV-01) was carried out in the UHF band and is centred on \(\alpha = 23{\rm h}\,30{\rm m}\), \(\delta = -55^\circ\). The survey is comprised of 78 pointings, each with a primary beam FWHM of $\sim$1.6~deg at 816~MHz. Each pointing was observed for $\sim$ 1 hour, achieving a per-pointing depth of 10~$\mu$Jy beam$^{-1}$ and brightness temperature sensitivity of $\sim$ 258 mK, for Briggs-weighted (\textsc{Robust} = $-$0.5) images. The combined mosaic has a sensitivity of 14~$\mu$Jy beam$^{-1}$ when smoothed to a common resolution of 10.2 arcsec. The total survey observation time was 116~hr and used an 8-second integration time in the 32k correlator mode setup, resulting in 32,768 channels, each 16.602 kHz wide. The combined data volume is approximately $\sim$250 TB. Out of 64 MeerKAT antennas, 60 were typically available for each pointing. The sources J0408$-$6545 (primary calibrator) and J2329$-$4730 (secondary calibrator) were used for absolute flux, bandpass, phase and complex gain calibration.

The MeerKAT data were calibrated and imaged using the same procedures as outlined in \citet{Magolego_2025}. In brief, the calibration was performed using the \textsc{Oxkat} pipeline \citep{Heywood_2020}, a set of \textsc{Python}-based scripts that semi-automatically process Stokes~I MeerKAT data. This pipeline performs both direction-independent and direction-dependent calibration, the latter being essential for our observations given MeerKAT’s large field-of-view, high sensitivity, and $\lesssim$1~GHz observing frequency. The workflow further includes point-source modeling and subtraction, followed by imaging strategies optimized for recovering diffuse, low-surface-brightness emission. We direct the reader to \citet{Magolego_2025} for full details of the data reduction and imaging pipeline.

We identified the diffuse source in Abell 4067 through a systematic manual search for extended radio emission in galaxy clusters within the MeerKAT-SPT 100~deg$^2$ field. The search encompassed >200 galaxy clusters drawn from two SPT cluster catalogues \citep{Huang_2020,Kornoelje_2025} overlapping the field. Although the source lies within the survey footprint, the host cluster is not part of the SPT cluster catalogues, but is instead identified in the X-ray-selected REFLEX~II cluster catalogue \citep{REFLEXII_2013,Xu_2022}.
\subsection{X-ray Data}
The X-ray image of Abell 4067 was produced from archival XMM-Newton EPIC-MOS observations (Obs.ID 0677180601, PI: Chon), with a total effective exposure time of 11 ks. We used data in the 0.5-2 keV band, which maximizes signal-to-noise for thermal intracluster medium (ICM) emission while minimizing background contamination.

Standard XMM-Newton data reduction was performed using the Science Analysis System (SAS), following recommended procedures for flare screening, event filtering, and background subtraction \citep[see][for more details]{Chon_2015}. Soft-proton flares were removed using the high-energy light-curve filtering, and only good events were retained (PATTERN $\lesssim$ 12 for MOS; FLAG == 0). Exposure maps were generated to correct for instrumental variations and vignetting. The final image was adaptively smoothed (12 arcsec) to enhance low-surface brightness features, allowing a clearer comparison between the ICM morphology and the radio emission.
\section{RESULTS}
\label{sec3}
\begin{figure*}
\centering
\includegraphics[width=0.9\textwidth, , trim=0 130 0 130,clip]{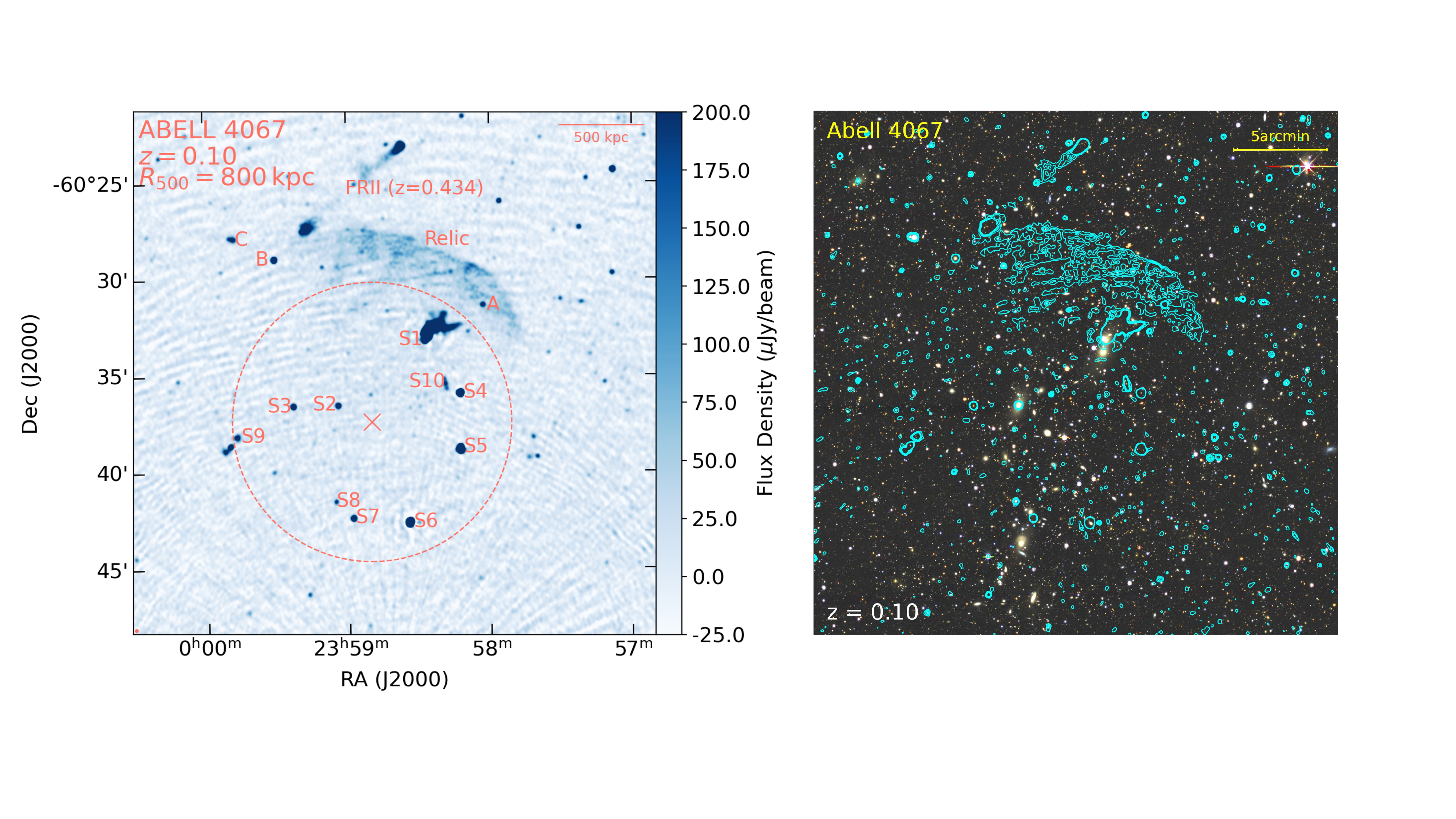}
\caption[Fullband_img]{Left: MeerKAT 816 MHz image of Abell 4067 at \textsc{Robust} 0.0, with a resolution of 11.7 arcsec $\times$ 11.7 arcsec, position angle 0.0 degree and rms noise, $\sigma$ = 6.20 $\mu$Jy beam$^{-1}$. The labels S1 - S10 indicate the compact sources within the $R_{500}$ cluster region (dashed orange circle) and those near the extended source labelled ‘Relic’ are A, B, C and FRII. The X-ray cluster centre is marked by ‘X’. The synthesized beam is shown in orange in the lower-left corner. Right: DESI Legacy Survey grz image of the region of the labelled sources with the MeerKAT 816 MHz contours overlaid. Contour levels are drawn at [5, 9, 13, 17] $\times$ 1$\sigma$, where $\sigma$ = 6.20 $\mu$Jy beam$^{-1}$.}
\label{RadioVsOptical}
\end{figure*}

The MeerKAT 816 MHz image is shown in Figure \ref{RadioVsOptical}, alongside an overlay on the Dark Energy Spectroscopic Instrument (DESI) Legacy Imaging Survey \citep{Dey_2019} grz-band image (Figure \ref{RadioVsOptical}). No central diffuse radio emission is found in our MeerKAT 816 MHz image. However, the field of Abell 4067 contains several discrete sources and a prominent elongated radio structure. The basic properties of the cluster are presented in Table \ref{properties} and the radio properties of the latter sources are detailed in the following subsections.

\subsection{Discrete Sources}
The discrete radio sources in the central region of the Abell 4067 field are designated S1-S10 (within the $R_{500}$\footnote{$R_{500}$ represents the radius within which the average density of the galaxy cluster is 500 times the critical density of the universe at the redshift of the cluster.} region) as depicted in Fig \ref{RadioVsOptical}, while those near the large elongated radio source are labeled A, B, C, and FRII. All sources in the S1-S10 group (including A-C and FRII) are resolved at 816 MHz and have optical counterparts in the DESI Legacy Imaging Survey Data (DR8; \citealt{Dey_2019,Duncan_2022}). Their properties are detailed in Table \ref{CompactSources}. Based on photometric redshifts, sources S1, S2, and S6 are identified as cluster members of Abell 4067, whereas S3, S4, S5, S7, S8, S9, and S10 are background galaxies. Source S1 is a possible head-tail radio galaxy located toward the cluster centre, identified as 6dFGS gJ235827.3$-$603311 \citep[$z~=~0.0978$;][]{Teague_1990}. Sources A, B, and the FRII radio galaxy \citep[e.g.,][]{Fanaroff_1974} are also background galaxies, source C, is identified as a foreground galaxy. However, source C has a photometric redshift consistent within the uncertainties to that of the cluster, suggesting it may be associated with Abell 4067.

Furthermore, source S2 is identified with the galaxy 2MASX J23590416$-$6036344 ($z = 0.100$; \citealt{Teague_1990}). Its optical, radio, and X-ray positions agree to within $\sim$1 arcsec, consistent with the astrometric uncertainties, confirming that they correspond to the same system. The X-ray feature is not emission from the galaxy itself, but rather from the cool, dense gas of the compact subcluster to which S2 belongs. This gas is being ram-pressure stripped as the subcluster penetrates the diffuse main cluster, producing the southwest-directed X-ray tail (see Figure~\ref{X_ray}). \citet{Chon_2015} showed that Abell~4067 consists of a merging pair: a small, cool-core component moving through a larger, low-density cluster, forming a bullet-like structure (e.g., \citealt{Sikhosana_2022}). The compact component corresponds to the S2 system.

\begin{table*}
\caption{The properties of the compact radio sources within and outside the $R_{500}$ cluster region of Abell 4067.}
\small 
\centering
\begin{tabular}{cccccc}
\toprule\toprule
\textbf{ID} & \textbf{RA (J2000)} & \textbf{Dec (J2000)} & \textbf{$z_{\mathrm{phot}}$} & \textbf{Flux} & \textbf{Notes} \\
\midrule
&(\text{$^{\mathrm{h}}$:$^{\mathrm{m}}$:$^{\mathrm{s}}$}) & (\text{$^{\circ}\!:\!^{\prime}\!:\!^{\prime\prime}$}) & & \text{(mJy)} &  \\
\midrule
S1 & 23:58:27.36 & $-$60:33:11.52 & 0.097737 & 61.2 $\pm$ 0.09 & Cluster member \\
S2 & 23:59:04.10 & $-$60:36:34.14 & 0.100659 & 0.72 $\pm$ 0.02 & Cluster member \\
S3 & 23:59:23.05 & $-$60:36:35.65 & 0.563 & 0.92 $\pm$ 0.02 & Background galaxy \\
S4 & 23:58:12.57 & $-$60:35:56.41 & 3.709 & 6.77 $\pm$ 0.02 & Background galaxy \\
S5 & 23:58:12.52 & $-$60:38:50.31 & 1.108 & 7.36 $\pm$ 0.03 & Background galaxy \\
S6 & 23:58:33.64 & $-$60:42:40.2 & 0.094 & 13.9 $\pm$ 0.03 & Cluster member \\
S7 & 23:58:58.05 & $-$60:42:23.42 & 1.847 & 1.14 $\pm$ 0.02 & Background galaxy \\
S8 & 23:59:05.45 & $-$60:41:31.8 & 0.181069 & 0.26 $\pm$ 0.02 & Background galaxy \\
S9 & 23:59:46.95 & $-$60:38:09.21 & 2.392 & 0.44 $\pm$ 0.01 & Background galaxy \\
S10 & 23:58:18.76 & $-$60:35:27.32 & 0.734 & 0.71 $\pm$ 0.03 & Background galaxy \\
A & 23:58:02.73 & $-$60:31:22.16 & 1.269 & 0.30 $\pm$ 0.01 & Background galaxy \\
B & 23:59:30.55 & $-$60:28:59.28 & 0.181 & 1.50 $\pm$ 0.02 & Background galaxy \\
C & 23:59:47.61 & $-$60:27:54.40 & 0.069188 & 0.74 $\pm$ 0.03 & Foreground galaxy \\
FRII (core) & 23:58:56.68 & $-$60:25:06.28 & 0.434 & 0.15 $\pm$ 0.01 & Background galaxy \\
\bottomrule
\end{tabular}
\label{CompactSources}
\end{table*}

\subsection{Radio Relic}
\begin{figure*}
\centering
\includegraphics[width=0.9\textwidth, , trim=0 220 100 180,clip]{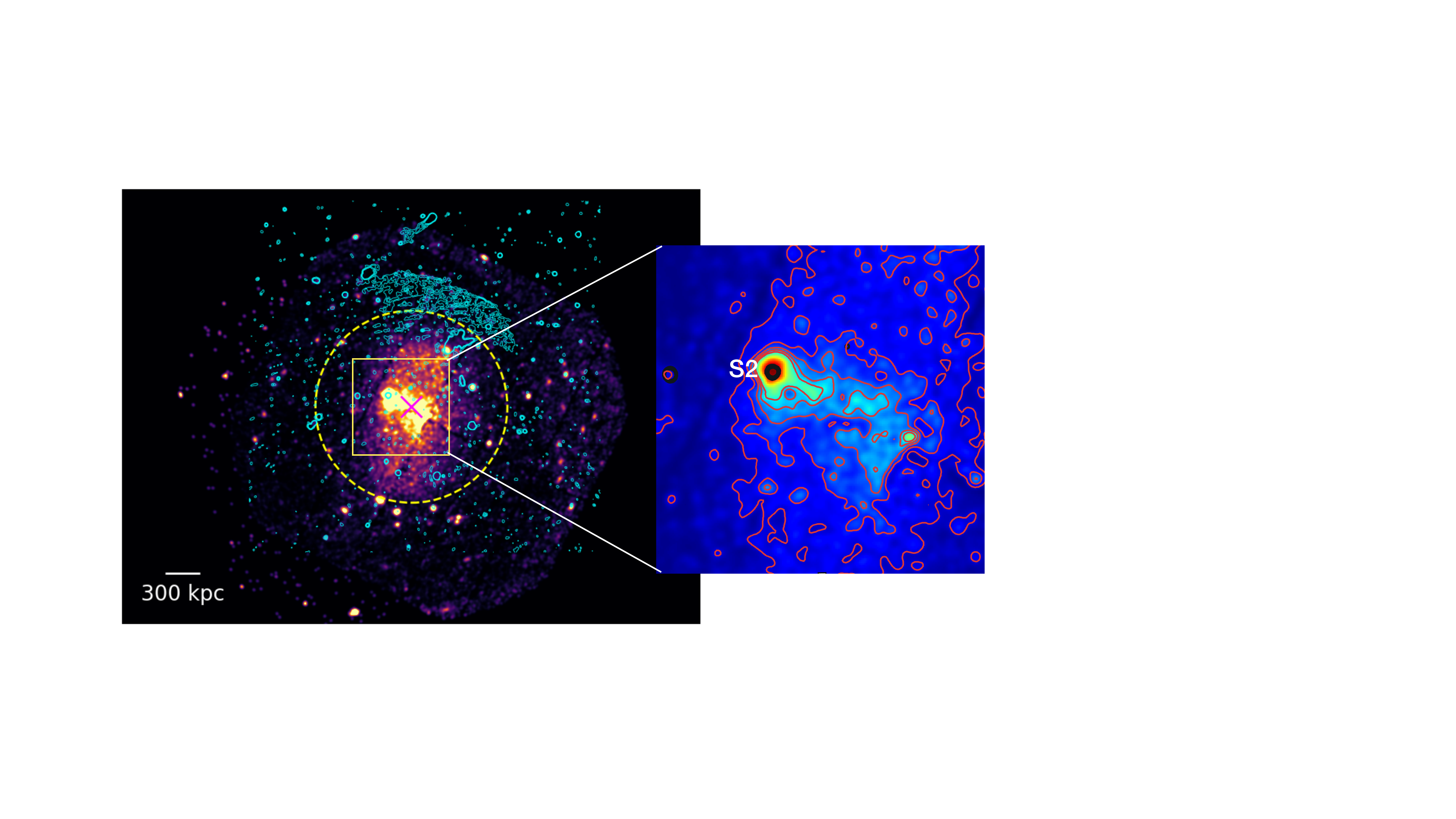}
\caption[UHF images]{Left: XMM-Newton X-ray image of Abell 4067 (smoothed to 12 arcsec) in the 0.5 - 2.0 keV band. The orientation of the radio relic is perpendicular to the major axis of the X-ray emission from the ICM. The symbol ‘X’ mark the position of the cluster according to the XMM-Newton and the yellow dashed circle indicate the $R_{500}$ radius. Right: Zoomed-in view of the region around the compact, bullet-like galaxy S2 (2MASX J23590416$-$6036344) as it penetrates the low-density outer ICM \citep{Chon_2015}. Red contour levels show the X-ray emission from the 12 arcsec smoothed XMM-Newton image, while black contour levels show the radio emission from the MeerKAT 816 MHz full-resolution image (8.9 arcsec $\times$ 8.9 arcsec, \textsc{Robust} $-0.5$).}
\label{X_ray}
\end{figure*}

We detect a diffuse, elongated radio source at a projected distance of approximately 954 kpc north of the X-ray centre of Abell 4067. This source has no optical counterpart in the DESI DR8 images (\citealt{Dey_2019,Duncan_2022}). We classify this structure as a radio relic. The source's position within the system, with its major axis oriented perpendicular to the primary elongation of the ICM thermal emission (which typically signifies the merger axis; see Figure \ref{X_ray}), its brightness decreasing toward the cluster centre, and the lack of a distinct optical counterpart, all support this classification. 

Assuming the relic is at the cluster redshift, its largest linear size (LLS) is $1.48 \pm 0.02$~Mpc. The LLS was calculated as the maximum separation between pixels identified as relic emission along the 5$\sigma$ contour, following \citet{Jones_2023}. The uncertainty corresponds to one synthesized beam, which represents the smallest angular scale over which the flux distribution can be reliably measured. The relic width was estimated statistically by drawing lines perpendicular to the LLS at each pixel along its length and measuring the maximum separation between relic pixels along those directions. We adopt the median of these measurements as the characteristic width ($320~\pm~21$~kpc), with the standard deviation as its uncertainty. Both the LLS and width should be regarded as lower limits, as the full extent of the relic emission may fall below the detection threshold. These uncertainties are compounded by limitations in the data: residuals from imperfect subtraction of both point-like and extended sources, as well as the creation of low-resolution images to enhance diffuse emission, can affect the measured sizes. 

Additionally, a strong source near the cluster centre introduces significant direction-dependent effects (DDEs), which we attempted to mitigate through peeling. However, residual calibration errors remain and vary across frequency. When the data are divided into sub-bands, the reduced signal-to-noise (SNR) ratio and more limited $uv$-coverage amplify these residual artefacts. Since spectral index measurements rely on small relative flux differences between frequencies, even modest frequency-dependent calibration errors can introduce artefacts in the in-band spectral index maps. In contrast, the integrated flux density measurements are derived from the full-band image, where the broader bandwidth improves SNR and stabilizes deconvolution. Frequency-dependent residual errors partially average out across the band, and the flux measurement depends primarily on the overall amplitude calibration rather than relative sub-band differences. Therefore, while the calibration limitations prevent reliable in-band spectral index estimates, their impact is less severe in the full-bandwidth map. Together, these factors limit the precision with which the relic structure can be characterised.
\subsection{Absence of Radio Halo}
Despite deep radio imaging ($\sim$ 10 $\mu$Jy/beam) and the low redshift of the cluster, no central radio halo is detected within the cluster, down to the sensitivity limits of our observations. The central region of the cluster, within $R_{500}$ = 800 kpc, shows no evidence of diffuse Mpc-scale emission. The non-detection of a halo is notable given the cluster's merging state, which has been confirmed through X-ray morphological parameters \citep{Chon_2015}. The cluster has an estimated total mass of $M_{500} = 2.0 \pm 0.25\times 10^{14}$ M$_{\odot}$, estimated from X-ray observations under the assumption of hydrostatic equilibrium and supported by temperature-mass scaling relations and galaxy velocity dispersion measurements, placing it among the lower-mass systems hosting radio relic.

The absence of a central halo may be intrinsic, reflecting a physical "off-state" in which the cluster simply lacks sufficient turbulent energy (see Section~\ref{lack}), or extrinsic, due to observational limits such as sensitivity and short-baseline coverage \citep{Bruno_2023}. 

\begin{figure}
\centering
\includegraphics[width=1\linewidth]{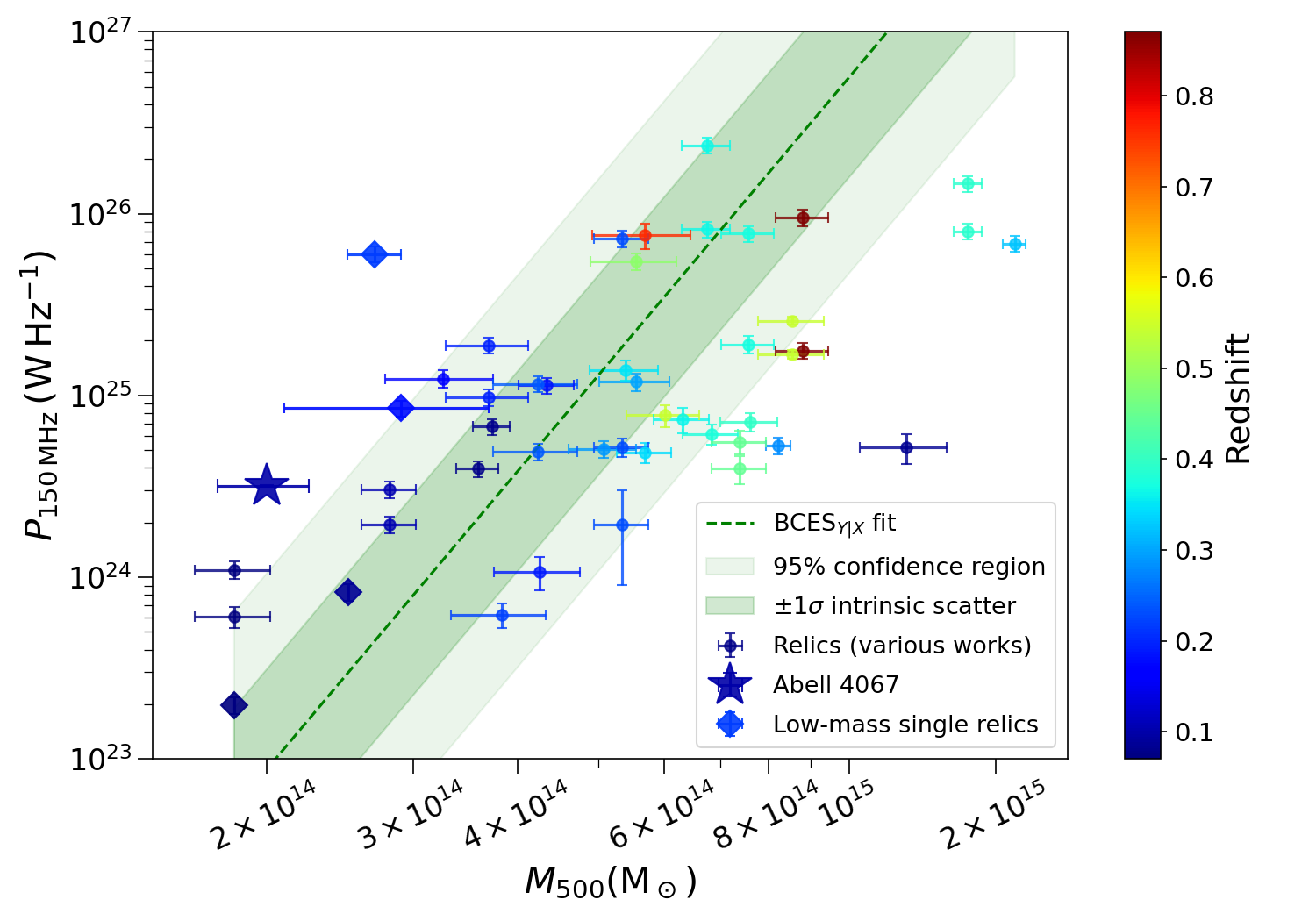}
\caption[Scaling relation]{The $P_{150\,\mathrm{MHz}}$ - $M_{500}$ scaling relation of radio relics, colour-coded by redshift. The solid line represents the best-fit relation, based on the correlation reported by \citet{Jones_2023}. The darker shaded region indicates the $\pm1\sigma$ intrinsic scatter (0.55 dex), while the lighter shaded region shows the corresponding 95$\%$ intrinsic scatter envelope. The radio relic in Abell~4067 is marked with a blue star and lies $\sim2.8\sigma$ above the best-fitting relation. Low-mass single relic systems are highlighted with blue diamonds.}
\label{ScalingRelation}
\end{figure}
\section{DISCUSSION}
\label{sec4}
\subsection{Detection and Morphology of the Radio Relic}
The MeerKAT UHF-band image reveals a radio relic in the outskirts of Abell 4067, with a largest linear size (LLS) of $1.48 \pm 0.02$ Mpc at 816~MHz. The relic exhibits an elongated arc-like morphology (see Figure \ref{RadioVsOptical}). Its orientation is distinctly perpendicular to the major axis of the X-ray emission from the intracluster medium (ICM), as shown in Figure \ref{X_ray}. Notably, this relic is the only one detected across the 100~deg$^2$ region of uniform MeerKAT sensitivity examined in our survey. Within this area, more than 200 SZ-selected galaxy clusters are present, implying an observed relic occurrence rate of $\sim$0.5 per cent in this uniform-sensitivity radio survey. This fraction is substantially lower than the relic incidence reported in targeted cluster surveys \citep[e.g., 53 percent;][]{Knowles_2022,Kolokythas_2025}, where systems are often pre-selected to be massive and dynamically disturbed. The low occurrence rate found here suggests that radio relics become increasingly rare in the general cluster population below masses of $3\times10^{14} M_{\odot}$. Given the relatively low mass of the host cluster, this suggests that low-mass systems may indeed contribute significantly to the relic population, provided one reaches sufficient surface-brightness. This interpretation is supported by the recent LoTSS-DR2-PSZ2 study, which finds that only $\sim$10 percent of Planck-selected clusters ($\sim$194) host radio relics at 144 MHz, but also suggests that many low-power relics remain undetected due to sensitivity limits \citep{Jones_2023}.
\subsection{Association with Merger Shock Waves}
The orientation of the radio relic and its projected distance ($\sim$~0.95~Mpc) from the cluster centre is consistent with the scenario in which radio relics trace merger shock waves (e.g., \citeauthor{Ensslin_1997} \citeyear{Ensslin_1997}). The relic spans a linear size of 1476 $\times$ 320 kpc at 816~MHz, with a measured flux density of $S_{816} =$ 27.5 $\pm$ 0.28 mJy. This measurement excludes the flux density contribution of source A (Gaussian fit), which is the only point source embedded within the relic. 
The flux densities and their uncertainties were calculated using the \textsc{Radioflux}\footnote{\url{https://github.com/mhardcastle/radioflux}} tool, which measures integrated fluxes by summing pixel values within user-defined source regions on the images.
\subsection{Spectral Considerations and Radio Power Estimation}
Due to residual calibration errors, we are unable to robustly estimate the integrated spectral index of the radio relic. We additionally attempted to construct a spectral index map using publicly available ASKAP EMU data \citep{Norris_2011,Hotan_2021}, but the image is dominated by artefacts to the extent that the relic is not detectable, preventing a meaningful comparison. Therefore, we assume a typical radio relic spectral index of $\alpha = 1.0$ (e.g., \citeauthor{Kale_2017} \citeyear{Kale_2017}; \citeauthor{Botteon_2022} \citeyear{Botteon_2022}). By extrapolating the flux density from 816 MHz to 150 MHz, we estimate a radio power of $3.10 \pm 0.03 \times 10^{24}$ W Hz$^{-1}$ at 150 MHz.
\subsection{Comparison with Known Radio Relics}

We compare the estimated radio power of the Abell~4067 relic to a sample of known radio relics \citep[$z \sim 0.07$-0.87;][]{deGasperin_2014,Locatelli_2020,Botteon_2022,Chatterjee_2022,Jones_2023}, including single relics hosted by low-mass clusters at $z \sim 0.1$ \citep{Kale_2017,Paul_2020,Paul_2021}. 
When plotting the $P_{150\,\mathrm{MHz}}$--$M_{500}$ relation, Abell~4067 is among the lowest-mass systems known to host a radio relic (Figure~\ref{ScalingRelation}). The best-fit line and uncertainty indicated by shading shown in Figure~\ref{ScalingRelation} is taken from the relic scaling analysis of \citet{Jones_2023}. The sample analysed by \citet{Jones_2023} is drawn from Planck SZ-selected clusters and is therefore dominated by high-mass systems. Of the 35 clusters in their study, 23 lie at $z \geq 0.2$ and typically have $M_{500} \sim 5-8 \times 10^{14}\,M_\odot$. The lower-mass regime is represented primarily by the 12 clusters at $z < 0.2$, which span $M_{500} \sim 1.8-4.3 \times 10^{14}\,M_\odot$, with a characteristic mass of $\sim 3 \times 10^{14}\,M_\odot$. Thus, while \citet{Jones_2023} provide robust constraints on relic occurrence in massive clusters, the statistics for lower-mass systems remain limited and largely confined to the nearby Universe.

The only lower-mass system in that work is PSZ2~G089.52+62.34 (A1904; \citeauthor{Weeren_2021} \citeyear{Weeren_2021}; \citeauthor{Paul_2021} \citeyear{Paul_2021}; \citeauthor{Botteon_2022} \citeyear{Botteon_2022}), which hosts two relics on the same side of the cluster and does not show a central radio halo. According to simulations by \citet{Nuza_2012}, the probability of detecting a radio relic in a cluster with mass $\sim 3.3\times10^{14}\,\mathrm{M}_\odot$ at $z\sim0.2$ is only a few percent with current radio facilities \citep{deGasperin_2014}. In this context, the detection of a relic in Abell~4067, with $M_{500}=2.0\times10^{14}\,\mathrm{M}_\odot$, further extends the known relic population towards the low-mass regime.

Figure~\ref{ScalingRelation} also shows that Abell~4067 lies $\sim 2.8\sigma$ above the BCES$_{Y|X}$ scaling relation, formally placing it beyond the 95$\%$ intrinsic scatter region. This indicates that the relic has comparatively high radio power for its mass, although it remains consistent with the broad intrinsic dispersion observed in relic scaling relations. Its position highlights that low-mass clusters can occasionally host relatively luminous relics, but its absolute radio power is still lower than that of relics in more massive systems.

Taken together, the position of Abell~4067 in the low-mass regime and the significantly lower radio power of relics hosted by low-mass clusters support a scenario in which merger shocks in less massive clusters are, on average, weaker and less efficient in accelerating relativistic particles \citep[e.g.,][]{Schellenberger_2022}. 
The colour-coded redshift distribution in Figure~\ref{ScalingRelation} further suggests that higher-redshift relics tend to be more luminous, which may partly reflect selection effects favouring the detection of intrinsically bright relics at larger distances, and/or evolutionary trends in merger-driven radio emission.

The discovery of a radio relic in such a low-mass system therefore provides an important observational constraint on relic formation models. Future deep MeerKAT observations and spatially resolved spectral studies, in concert with other major radio facilities, will be essential for constraining the shock properties and particle-acceleration efficiency in low-mass mergers and for establishing the statistical behaviour of relics across a wider cluster mass range.

\subsection{The Lack of a Radio Halo}
\label{lack}
Despite clear merger indicators and shock evidence (also implied by the radio relic morphology), no diffuse central radio halo is observed. This absence is notable in light of the established correlation between cluster mergers and halo occurrence \citep[e.g.,][]{Brunetti_2014,Cuciti_2015}. Several factors likely contribute:
\begin{enumerate}[label=\roman*), leftmargin=*, topsep=2pt, itemsep=1pt]
    \item Low Cluster Mass: With $M_{500} = 2.0 \times 10^{14}$ M$_{\odot}$, this cluster lies below the empirical threshold $M_{500} \sim 6 \times 10^{14}$ M$_{\odot}$ above which radio halos are more frequently detected \citep{Cuciti_2015}. This supports the idea that lower-mass systems may lack the turbulent energy required to re-accelerate relativistic electrons across cluster cores \citep{Cassano_2013,Cuciti_2021a,Cassano_2023}.
    \item Merger Stage: The absence of a halo might reflect a late-stage merger. If turbulence has already dissipated, it would no longer power a central halo \citep[e.g.,][]{Brunetti_2009,Bonafede_2017}. Turbulent energy in the ICM typically dissipates on timescales of $\sim$0.5 - 1 Gyr \citep{Donnert_2013,Brunetti_2014,Weeren_2019}, after which the relativistic electrons cool and the halo fades. The presence of a surviving radio relic supports this timing, as relics may outlive halo emission due to their localized, shock-induced nature.
    \item Shock vs. Turbulence: While shocks can efficiently re-accelerate fossil plasma to produce relics and phoenices, the lack of sustained, volume-filling turbulence could prevent the formation of a halo \citep[e.g.,][]{Cassano_2006,Cassano_2009,Brunetti_2011,Hallman_2011}.
\end{enumerate}
Altogether, Abell 4067 highlights the importance of studying mergers in the low-mass regime, where relic-no-halo systems may be more common than currently appreciated \citep[e.g.,][]{Bonafede_2017,Weeren_2021,Paul_2021,Botteon_2022,Jones_2023}. Expanding the number of well-characterised clusters below $M_{500} \lesssim 3 \times 10^{14}$ M$_\odot$ \citep{Bleem_2024,Kornoelje_2025} is essential for establishing how merger energetics, shock strength, and turbulence generation scale with mass. In this context, the growing MeerKAT-SPT survey designed to systematically probe low- and intermediate-mass clusters with high radio sensitivity will provide the larger statistical samples needed to clarify the conditions under which relics can form in the absence of radio halos and to refine models of particle re-acceleration in the intracluster medium.
\section{Conclusion}
\label{sec5}
In this Letter, we report the discovery of a single arc-like radio relic in the galaxy cluster Abell~4067, one of the lowest-mass clusters known to host a single arc-like relic. The MeerKAT 816~MHz image reveals a relic with a largest linear size of $\sim 1.48 \pm 0.02$~Mpc, located at a projected distance of $\sim$0.95~Mpc from the cluster centre. XMM-Newton X-ray observations show that both the location and orientation of the relic relative to the elongation of the intracluster medium are consistent with a scenario in which the relic traces a merger-driven shock propagating in the cluster outskirts. We measure the radio power of the relic to be $3.10 \pm 0.03\times10^{24}$~W~Hz$^{-1}$ at 150~MHz,  assuming a spectral index $\alpha = 1$. 

When compared with the population of known radio relics, Abell~4067 occupies the extreme low-mass regime of relic-hosting clusters. Its relic lies relatively high in radio power for its mass, ~2.8$\sigma$ above the BCES${Y|X}$ scaling relation, yet its absolute radio power remains lower than that typically observed in more massive systems. This highlights that low-mass clusters can occasionally host comparatively luminous relics, but the large intrinsic scatter in the $P_{150\,\mathrm{MHz}}$–$M_{500}$ relation prevents drawing strong conclusions about mass dependence from a single system.

Taken together, the low cluster mass of Abell~4067 and the significantly lower radio-power distribution of relics in low-mass systems are consistent with a scenario in which merger shocks in less massive clusters are, on average, weaker and less efficient at accelerating relativistic particles, resulting in reduced synchrotron luminosity.

The absence of a detectable radio halo in Abell~4067, despite clear evidence for an ongoing merger, could point to a dependence of radio halo formation on cluster mass, merger energetics and evolutionary stage. This system therefore provides a valuable testbed for models of particle re-acceleration and turbulence in the intracluster medium, and reinforces the view that radio halos are not a universal outcome of cluster mergers.

Future deeper MeerKAT observations and spatially resolved spectral studies will enable tighter constraints on the shock properties and particle-acceleration efficiency in this relic, and will place Abell~4067 in a broader statistical context as the sample of low-mass relic-hosting clusters continues to grow.

\section*{Acknowledgements}

ISM and RPD acknowledge the financial support of the South African Radio Astronomy Observatory (SARAO) for this research. The MeerKAT telescope is operated by SARAO, a facility of the National Research Foundation (NRF), which is an agency under the Department of Science, Technology and Innovation (DSTI). We also express our gratitude to the SARAO science commissioning and operations team, led by Sharmila Goedhart, for their assistance. We also acknowledge the use of the ilifu cloud computing facility (\url{www.ilifu.ac.za}), a partnership between the University of Cape Town, the University of the Western Cape, Stellenbosch University, Sol Plaatje University, and the Cape Peninsula University of Technology. The ilifu facility is supported by contributions from the Inter-University Institute for Data Intensive Astronomy (IDIA), a partnership between the University of Cape Town, the University of Pretoria, and the University of the Western Cape; the Computational Biology division at UCT; and the Data Intensive Research Initiative of South Africa (DIRISA). MA is supported by FONDECYT grant number 1252054, and gratefully acknowledges support from ANID Basal Project FB210003,  ANID MILENIO NCN2024$\_$112 and ANID + Vinculaci\'on Internacional + FOVI250261. Argonne National Laboratory’s work was supported by the U.S. Department of Energy, Office of High Energy Physics, under contract DE-AC02-06CH11357.\\\\
Additionally, this work made use of the CARTA (Cube Analysis and Rendering Tool for Astronomy) software (DOI: \url{10.5281/zenodo.3377984} – \url{https://cartavis.github.io}).

\section*{Data Availability}

The MeerKAT raw data used in this study are publicly available (Project IDs SCI-20220822-JV-02 and SCI-20230907-JV-01) in accordance with the South African Radio Astronomy Observatory data release policy, while calibrated and advanced data products are available upon reasonable request.



\bibliographystyle{mnras}
\bibliography{example} 








\bsp	
\label{lastpage}
\end{document}